\documentstyle[preprint,aps]{revtex}
\tightenlines
\begin{document}
\draft
\preprint{SUNY-FRE-98-09}
\title{Gauge Invariant SO(3) - Z2 
Monopoles as Possible Source of Confinement in SU(2)
Lattice Gauge Theory}
\author{Michael Grady}
\address{Department of Physics, SUNY College at Fredonia, 
Fredonia NY 14063 USA}
\date{\today}
\maketitle
\begin{abstract}
A gauge invariant procedure for extracting combined SO(3)-Z2 
monopoles in positive-plaquette SU(2) lattice gauge 
theory is shown. When these monopoles are eliminated 
through a constraint, the theory deconfines for 
all $\beta$ on $12^4$ and $20^4$ lattices even in the strong 
coupling limit, despite a rather strong average plaquette of 
around 0.64. This corresponds to an effective Wilson $\beta$ 
of 2.45, at which Wilson-action lattices would be far into the 
confining region.  This suggests that Wilson-action confinement 
may be a strong-coupling lattice artifact; 
the continuum limit may not confine.
\end{abstract}
\narrowtext

Rather strong evidence has been presented that 
confinement in SU(2) and SU(3) lattice gauge theories
is connected to the presence of long abelian monopole
current loops in the maximal abelian gauge\cite{mag}.  An alternative
hypothesis that it is the center of the gauge group which is 
important, center vortices in the maximal center gauge, also
seems to work\cite{greensite}. It has long been suspected 
that a gauge invariant monopole of some sort 
which is ultimately responsible for confinement
underlay such
objects only visible in certain gauges.
Following, a candidate for 
a gauge invariant monopole in SU(2) lattice gauge theory is presented.
It is an SO(3) monopole defined using the three bent double-plaquettes
which comprise a formulation of
the non-abelian Bianchi identity.  It is at the same
time a Z2 monopole. These monopoles are not in a 1-1 correspondence
with abelian monopoles in the maximal abelian gauge - at weak
couplings they are much more numerous - but there is 
an apparent connection.  When a constraint
prohibiting SO(3)-Z2 monopoles is added, 99.4\% of
abelian monopoles are eliminated when compared to the unconstrained
simulation, or 95\% when compared to a Wilson-action simulation at
the same renormalized coupling (matching average-plaquette values).
Thus the correlation between the two types of monopoles 
is quite strong,
in the sense that the vast majority of abelian monopoles are
eliminated when the SO(3)-Z2 monopoles are eliminated.
More importantly,
this constraint also results in deconfinement for all values 
of $\beta$.
Unlike previous restricted-action studies that also did not show
confinement\cite{loop}, 
the average-plaquette value in the $\beta \rightarrow 0$ limit here
is around 0.640 which lies in a region where 
the theory should be confining.  
The SO(3)-Z2 monopoles are lattice artifacts
in the sense that they require 
large-angle plaquettes to support them, so
they will not exist in the 
continuum. An action that prohibits them should
fall in the same universality 
class as the Wilson action, yet it does not
appear to confine. This severe lack of universality suggests 
that confinement in the ordinary Wilson theory may be
a strong-coupling lattice artifact, similar to the U(1) case.
    
We start with Wilson-action SU(2) lattice gauge theory with a 
positive-plaquette constraint, the positive-plaquette model\cite{pp}.
This constraint eliminates Z2 strings (strings of 
negative plaquettes).
Z2 strings are responsible for confinement in Z2 lattice gauge theory,
so eliminating them causes the Z2 theory to deconfine.
The positive-plaquette SU(2) model, however, 
still confines at small $\beta$ \cite{hellerpp}. 
So in SU(2) there must be something besides Z2 strings that causes
confinement. Actually Z2 strings probably {\em are}
responsible for 
confinement in the mixed fundamental-adjoint \cite{bhanot} version of
SU(2) in the large $\beta _A$ region, which includes the
Z2 theory as a limiting case. Because Z2 strings can cause confinement
(though they are not the only cause) a positive plaquette constraint
must be maintained along with any monopole constraint 
in order to get a possibly non-confined theory.

The identification of the monopole starts with
the non-abelian Bianchi identity\cite{bal,skala}.
This can be expressed by first constructing 
the covariant (untraced) plaquettes
that comprise the six faces of an elementary cube, with the necessary
``tails'' to bring them to the same starting site (Fig.~1). Call these
$A$, $B$, $C$, $D$, $E$, and $F$. Now construct three bent double 
plaquettes also shown in Fig.~1, $X=AB$, $Y=CD$, and $Z=EF$.  
If one forms the
product $XYZ$, each link will cancel with its conjugate, so
$XYZ=1$. This is the non-abelian Bianchi identity. 
Although the plaquettes
are all positive (due to the positive plaquette
constraint), and thus have a trivial Z2 component of unity, the
double plaquettes may be negative. Factor each of 
these into  Z2 and SO(3) (positive-trace) factors, e.g. 
$X=Z_{X}X'$ etc.,
where $Z_{X}=\pm 1$ and $\rm{tr} X' > 0$.  Then the Bianchi identity
reads $X' Y' Z' Z_{X} Z_{Y} Z_{Z}=1$. This can be realized in either 
a topologically
trivial or nontrivial way as far as the SO(3) group is concerned.
If $Z_{X} Z_{Y} Z_{Z}=1$ then $X'Y'Z'=1$. However if 
$Z_{X} Z_{Y} Z_{Z}=-1$ then
$X'Y'Z'=-1$. In this case one has an SO(3) monopole which,
since it also carries a Z2 charge, can be pictured to be at the same
time a Z2 monopole.   The decomposition of the double-plaquettes
into Z2 and SO(3) factors is gauge invariant
since the trace is invariant. 
In such a monopole a large SO(3) flux is
in some sense cancelled by a large Z2 
flux in order to satisfy the SU(2)
Bianchi identity.  This is reminiscent of the abelian monopole
in U(1), in which a large flux of $2\pi$ enters or 
exits an elementary cube.
This apparent non-conservation of flux 
is allowed by the compact Bianchi 
identity since $\exp (2\pi i)=1$.
In the continuum the Bianchi identity enforces 
exact flux conservation.
If plaquettes in U(1) are restricted 
to $\cos(\theta _p) > 0.5$, then
the only solution to the 
Bianchi identity is the topologically trivial one,
$\theta_{\rm{tot}}=0$,
where $\theta _{\rm{tot}}$ is the sum 
of the six plaquette angles in an
elementary cube. This eliminates the monopoles,
and shows that they 
are strong-coupling lattice artifacts. 
The U(1) lattice gauge theory,
as a result, is deconfined in the continuum limit.

The SO(3)-Z2 monopoles described above 
are also lattice artifacts.
If plaquettes are restricted 
so that $\cos (\theta _{p}) \! > \! \sqrt{2}/2$, 
then even the double-plaquettes are positive,
and the SO(3) monopoles described above cannot exist.  Since in
the continuum limit all plaquettes are in the neighborhood 
of the identity,
such a restriction should have no effect on the continuum limit.
Therefore, these monopoles will not exist in the continuum, and 
exact SO(3) flux conservation on elementary cubes will hold there.  
Indeed it
has long been recognized that if SU(2) confinement is due 
to monopoles or vortices, then the only
such objects which could survive the 
continuum limit to produce 
confinement there are large objects (fat monopoles
and vortices) for which flux
is built up gradually\cite{fat}.
A good way to look
for fat-monopole confining configurations would 
seem to be 
to choose an action
which eliminates the single lattice spacing
scale artifacts while still allowing
similar larger objects to exist. 

The plaquette constraint mentioned above is one such possibility, 
however this results in a rather weak renormalized coupling. 
There are other	less severe
constraints which will do. 
For instance,
one can prohibit the double plaquettes $X$, $Y$ and $Z$ from 
being negative.
This also has to be done for the other ordering for which
a non-equivalent set of double plaquettes can be 
defined, $\tilde{X}=BC$,
$\tilde{Y}=DE$, and $\tilde{Z}=FA$, which also results in a 
Bianchi identity, $\tilde{X}\tilde{Y}\tilde{Z}=1$. 
(In addition, the 
positive-plaquette
restriction is also applied.)
This results in a clearly deconfined
theory for all $\beta$ on a $12^4$ lattice, and eliminates 
all but a very
few stray abelian monopoles in the maximal abelian gauge; 
about one monopole per 
every 45 $12^4$ lattices
remains.  However
the average plaquette for the strongest $\beta$ 
simulated, $\beta=0.01$, is
about 0.715, which represents a fairly weak renormalized 
coupling,
corresponding to a Wilson-$\beta$ of about 2.9. At this $\beta$, 
the $12^4$
lattice is not expected to be in the confining region 
anyway - it
would be beyond the finite-temperature deconfining transition for this
lattice, which occurs around 2.62. 
Nevertheless, the number of 
abelian monopoles 
in a Wilson-action simulation at $\beta=2.9$ is
much larger, around 33
per $12^4$ lattice, 1500 times as many, so there
is certainly a non-universality of abelian 
monopole density as a function of
average plaquette, which can be thought of as the 
renormalized coupling
(or closely related to it).

An even more interesting action is to explicitly 
prohibit the SO(3)-Z2 monopoles from
forming in the update (again, both orderings must 
be considered and the positive plaquette constraint is applied).
This also results in a deconfined 
theory for all $\beta$ on
$12^4$ and $20^4$ lattices. Fig.~2 shows the average 
Polyakov loop modulus
$<\! |L|\! >$ vs. $\beta$, and Fig.~3 shows histograms of the 
Polyakov loop modulus
at $\beta=0.01$. Here the average plaquette is 
only 0.640, corresponding 
to an effective Wilson $\beta$ of 2.45,  
well into the strong-coupling side of
the crossover
region where these lattices would be 
clearly confined in a Wilson-action
simulation.  Also, compared to the 
Wilson-action simulation at $\beta=2.45$,
there are only about 5\% as many abelian monopoles in the 
maximal abelian gauge. 
This is more than in the positive double-plaquette 
action, but apparently not
enough to confine. In this case the lack of universality
is clearly severe, with one action confining and the other not at
the same renormalized coupling.  Of course, only the weak coupling 
behavior of theories with different actions is universal. The strong
coupling behavior can be drastically different, due to, for example,
a phase transition.  This suggests that the universal continuum limit
of SU(2) lattice gauge theory is not confining, 
with confinement in the
Wilson-action theory at strong coupling being due 
to lattice artifacts, specifically
SO(3)-Z2 monopoles.
This confining phase would be separated from the continuum phase by 
a {\em zero-temperature} phase transition,
as in the U(1) case.  In other words, what is usually interpreted as
a finite-temperature deconfining transition could possibly be a
zero temperature transition, with $\beta _c$ 
dependent on lattice size,
but approaching a constant 
(perhaps around 2.8 \cite{wloop})  
rather than infinity in the infinite lattice limit.
 
It is interesting to consider whether the abelian 
monopoles remaining in
the SO(3)-Z2 monopoleless theory could result in confinement on 
much larger
lattices. This question 
can be explored with some confidence by looking
at the abelian monopole loop distribution function, which 
appears rather independent
of lattice size for monopole loops less 
than about 1.5 times the lattice size\cite{loop,wloop}.
The loop distribution function $P(l)$, is shown in Fig.~4
as a function
of loop length $l$. Here $P(l)$ is defined to be 
the probability, normalized
per lattice site, for a lattice to have a loop of 
length $l$.
As with the Wilson action\cite{wloop,teper} $P(l)$ appears 
to follow a power law, $P(l) \sim l^{-q}$ 
(except for loops of the minimal length 4).  
For the SO(3)-Z2
monopoleless action, a fit to Fig.~4 gives $q=5.2\pm 0.2$
for the $12^4$ lattice and also $5.2\pm 0.2$ for the $20^4$ lattice.
Size six loops, which fall slightly below the trend, were
excluded from the fits as were size four.
For abelian monopoles to cause confinement,
loops of at least length $N$, the lattice size, 
must exist (on most confining
lattices there are loops that wrap 
through the periodic boundary).
In fact, usually confinement does not set in 
until loops of length around
$N^2$ are common, due to loop crumpling.
The probability of a loop of size $N$ {\em or greater} 
existing on an $N^4$ 
lattice is given by 
\[ N^4 \sum^{\infty}_{l=N} P(l)  .
\]
Its behavior in the infinite lattice limit depends
on $q$. If $q > 5$,  then the above probability 
vanishes as $N \! \rightarrow \! \infty$ \cite{loop,wloop}. 
To eliminate any doubts concerning the value of $q$, a
simulation was performed at $\beta=0.1$ on the $12^4$ lattice.
Here the average plaquette was 0.653 corresponding to an effective
Wilson $\beta$ of 2.51, still fairly strong. The 
loop distribution was even steeper, with $q=6.0 \pm 0.4$.
Another run with $\beta = 0.5$ gave $q=8.2 \pm 0.6$.
From this it appears that not enough abelian monopoles
remain in the SO(3)-Z2 monopoleless theory to confine 
for any $\beta$ on any
size lattice.  
Indeed, since the restriction only affects 
single lattice-spacing scale objects, 
and the renormalized coupling is fairly
strong, it would seem that the any confining ``fat monopoles''
should have shown up
by the time a $20^4$ lattice was reached.

Although confinement was not seen on the lattices studied,
it is still conceivable that it could reappear on
very large lattices via some other mechanism, 
perhaps one
not associated with abelian monopoles. Another unlikely 
possibility is that the monopole constraint (and the positive
double-plaquette constraint that behaves similarly) somehow
erects a barrier to efficient Monte Carlo equilibration, and
that the lattices studied are all in some kind of
metastable state.
Most lattices studied were equilibrated for 500 sweeps after
which 500-6000 measurement sweeps were taken. There were no 
observable differences between earlier and later portions.
The maximum size of update was also varied, from a
completely open choice of new link, to one 
restricted to various-sized neighborhoods 
of the current 
link. There was no significant difference between these runs.
The Polyakov loop symmetry is so strongly broken in the SO(3)-Z2
monopoleless theory, even at $\beta=0.01$, 
that, although zero and slightly negative values
were fairly common, only one definite long-lived
tunneling to large negative Polyakov loop values 
was observed. 
It will likely take much longer runs or a non-local update scheme
to study tunneling in this theory. Symmetry breaking 
at larger $\beta$,
judged from the Polyakov loop value, is
even stronger.

It is interesting to speculate the 
mechanism by which SO(3)-Z2 monopoles
could cause confinement. The large violation of flux 
conservation that these entail could easily randomize the values
of Wilson loops.
Although the number of such monopoles decreases
as $\beta$ is increased, there are still a substantial number in
the deconfined region of Wilson-action simulations.
Fig.~5 shows the density of SO(3)-Z2 monopoles on $8^4$
and $12^4$ 
lattices 
with the standard Wilson action. The $8^4$ lattice deconfines around 
$\beta=2.4$, where no particular signal is evident in the 
monopole density. The situation 
here could be similar to that in the 
two-dimensional X-Y model, where it is not the number of vortices that
changes at the phase transition, but rather the binding 
of vortices into pairs.
Perhaps in the deconfined region most SO(3)-Z2 monopoles are in
closely bound pairs. Their flux violations could cancel locally and
not have much effect on Wilson loops. 
Another observation is that the monopole density falls below 1/4
at around $\beta=2.74$, fairly close to the point at
which $q$ becomes equal to 5 for the Wilson-action 
theory\cite{wloop}, where it is argued above that infinite-lattice
deconfinement must occur. It is possible that this point is
coincident with
a monopole occupation fraction of 25\%.

A gauge invariant SO(3) monopole, which is at the same time a Z2
monopole has been defined and proposed as the source of confinement
in SU(2) lattice gauge theory. Eliminating such monopoles
with a constraint results in a deconfined theory for all
couplings and lattice sizes, despite a rather strong renormalized
coupling. Since the constraint only affects the action at
strong coupling, universality
would seem to imply that all SU(2) lattice gauge theories are
non-confining in the continuum limit. This has been suggested
previously from other considerations\cite{zp,ps2,ggm}.
Another possibility could be that the SO(3)-Z2 monopoleless 
theory and the standard Wilson theory are in different universality
classes, however this would require a complete revision of
the usual understanding of the continuum limit.
A similar monopole
based on the Z3 center 
can be defined in
SU(3) lattice gauge theory. It would be interesting to
see if it is responsible for confinement there.  
An even more interesting question is how
the monopoleless theory behaves when light dynamical quarks are
added. If chiral 
symmetry still breaks spontaneously,
it is possible that chiral symmetry breaking will itself produce
an effective confining potential\cite{zp,nc}. 
Polarization effects in the 
chiral condensate could produce at least a partially 
chiral-expelled bag around hadrons which could result in a confining
potential (out to where string breaking by quark
pair creation occurs). Simulations showing
a reduced value of $<\! \bar{\psi} \psi \! >$
in the neighborhood of color sources have been reported\cite{markum},
in concert with this proposed mechanism, which has some features
in common with chiral quark models\cite{cqm}. Instantons may also 
play a role in confinement\cite{ilm}.  With the 
strong background of lattice artifacts removed, 
the SO(3)-Z2 monopoleless
action may be ideal for the study of lattice instantons.

\newpage
\begin{center}
{\Large Figure Captions}
\end{center}
\noindent
FIG. 1. Gauge covariant plaquettes and double plaquettes
from which non-abelian Bianchi identity is formed.
\newline
\noindent
FIG. 2. Polyakov loop modulus as a function of $\beta$ for the 
$12^4$ lattice with SO(3)-Z2 monopoleless action. 
\newline
\noindent
FIG. 3. Polyakov loop histograms for the $12^4$ 
(left scale) 
and $20^4$ (right scale) lattices at 
$\beta =0.01$.
\newline
\noindent
FIG. 4. Log-log plots of loop probability vs. loop
length for the SO(3)-Z2 monopoleless action, with linear fits. 
\newline
\noindent
FIG. 5. Density (number per link)
of SO(3)-Z2 monopoles vs. $\beta$
using the standard Wilson action.  \\\\

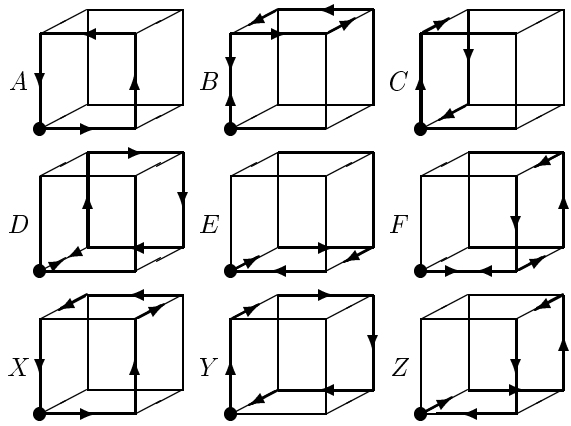
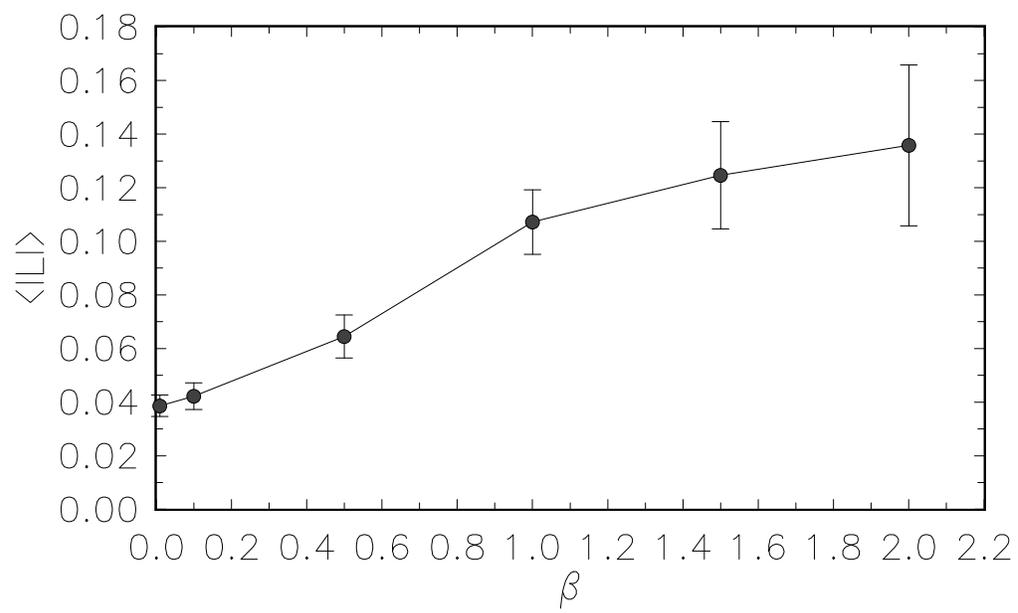
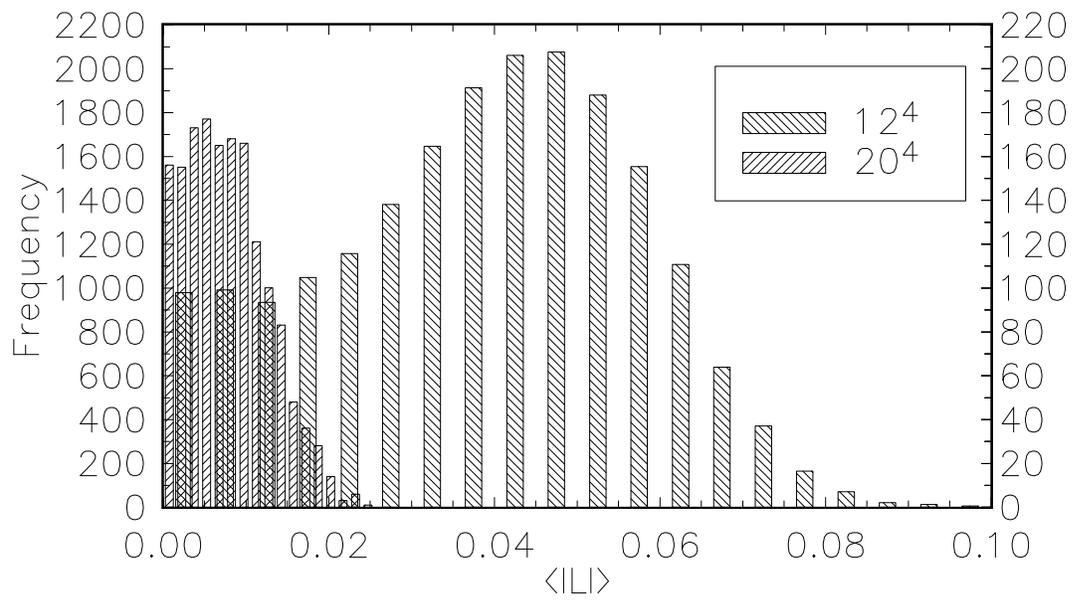
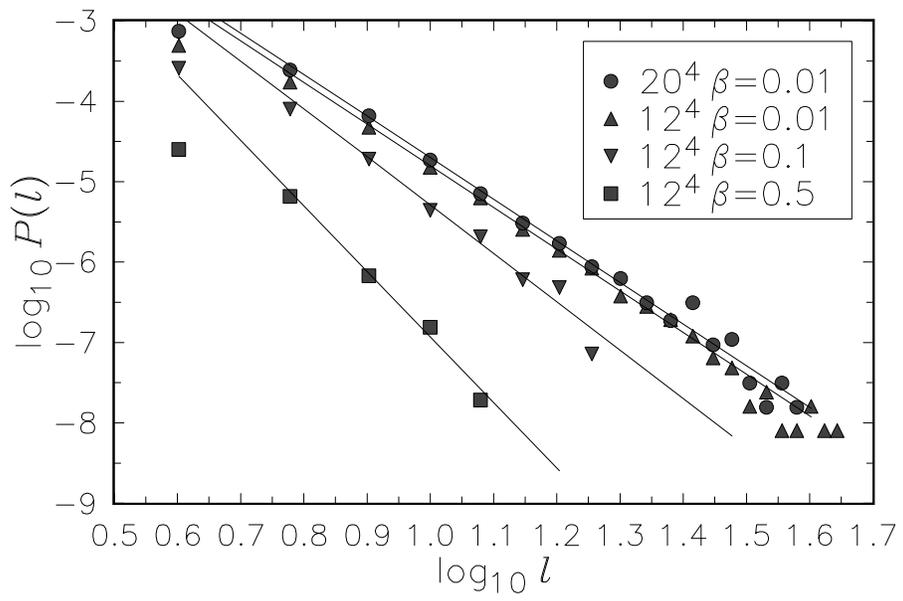
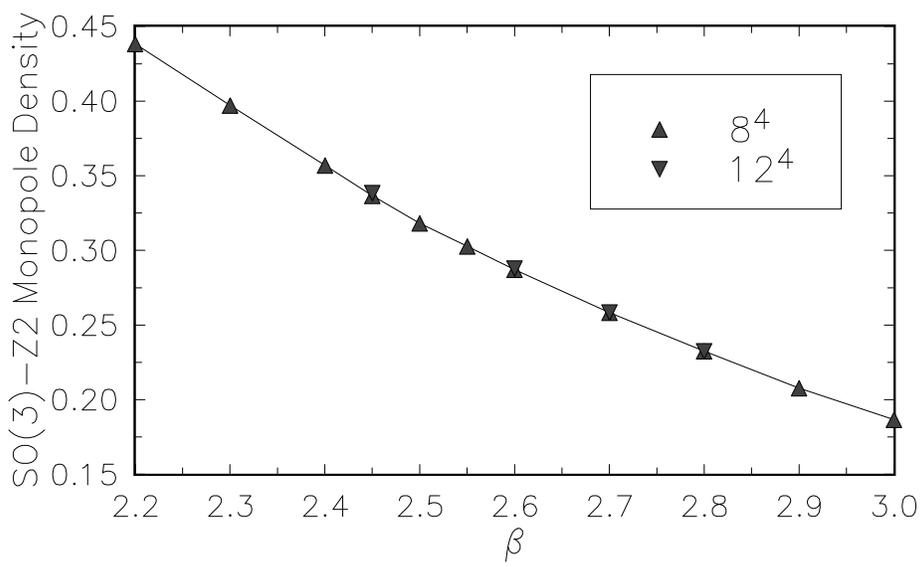
\end{document}